\documentclass[preprint,12pt]{elsarticle}

\usepackage[left=1in,right=1in,top=1in,bottom=1in]{geometry}
\usepackage{graphicx}
\usepackage{epstopdf}
\usepackage{amsmath}
\usepackage{mathtools}
\usepackage{amssymb}
\usepackage{upgreek}
\usepackage{float}
\usepackage{contour}

\usepackage{subcaption}
\usepackage[table]{xcolor}
\usepackage[normalem]{ulem}
\graphicspath{{./figures/},{./figures_eps/}}
\usepackage{tikz}

\usepackage{verbatim}
\usepackage{ulem}
\usepackage{color,soul}
\usepackage{enumerate}
\usepackage{enumitem}

\usepackage{natbib}
\usepackage[bookmarksnumbered=true,breaklinks=true,colorlinks,citecolor=blue,linkcolor=blue]{hyperref}
\setcitestyle{numbers,square}

\newcommand{\AL}[1]{\textcolor{orange}{*** #1 ***}}

\newcommand{\MDG}[1]{\textcolor{cyan}{*** #1 ***}}
\definecolor{ao(english)}{rgb}{0.0, 0.5, 0.0}

\newcommand{\KZrevise}[1]{\textcolor{black}{#1}}
\newcommand{\MDGrevise}[1]{\textcolor{black}{#1}}
\newcommand{\ALrevise}[1]{\textcolor{black}{#1}}

\newcommand\Rey{\mbox{\textit{Re}}}  

\begin{document}

\title{Turbulence control in plane Couette flow using low-dimensional neural ODE-based models and deep reinforcement learning}

\author[add1]{Alec J. Linot}
\author[add1]{Kevin Zeng}
\author[add1]{Michael D. Graham\footnote{Corresponding author: mdgraham@wisc.edu}}

\address[add1]{Department of Chemical and Biological Engineering, University of Wisconsin-Madison, Madison WI 53706, USA}

\date{\today}

\begin{abstract} 

The high dimensionality and complex dynamics of turbulent flows remain an obstacle to the discovery and implementation of control strategies. Deep reinforcement learning (RL) is a promising avenue for overcoming these obstacles, but requires a training phase in which the RL agent iteratively interacts with the flow environment to learn a control policy, which can be prohibitively expensive when the environment involves slow experiments or large-scale simulations. We overcome this challenge using a framework we call ``DManD-RL" (data-driven manifold dynamics-RL), which  generates a data-driven low-dimensional model of our system that we use for RL training. With this approach, we seek to minimize drag in a direct numerical simulation (DNS) of a turbulent minimal flow unit of plane Couette flow at $\Rey=400$ using two slot jets on one wall. We obtain, from DNS data with $\mathcal{O}(10^5)$ degrees of freedom, a $25$-dimensional DManD model of the dynamics by combining an autoencoder and neural ordinary differential equation. Using this model as the environment, we train an RL control agent, yielding a $440$-fold speedup over training on the DNS, with equivalent control performance. The agent learns a policy that laminarizes 84\% of unseen DNS test trajectories within $900$ time units, significantly outperforming classical opposition control (58\%), despite the actuation authority being much more restricted. The agent often achieves laminarization through a counterintuitive strategy that drives the formation of two low-speed streaks, with a spanwise wavelength that is too small to be self-sustaining. The agent demonstrates the same performance when we limit observations to wall shear rate.

\end{abstract}

\maketitle

\section{Introduction} \label{sec:Introduction}

Energy loss due to turbulent drag is ubiquitous in many industrial and commercial processes, ranging from air flowing over a plane wing, a ship in the ocean, or oil pumped through a pipe. In total, turbulent drag accounts for 25\% of the energy used in industry and commerce, resulting in 5\% of all man-made CO$_2$ emissions \cite{Jimenez2018a}. Even small reductions in this drag can yield massive savings in energy, which has long motivated the search for better flow control strategies.

Many control types for reducing turbulent drag exist, including, but not limited to, polymer/surfactant drag reduction \citep{Virk1975,Graham2014}, riblets \citep{Choi1989}, wall oscillations \citep{quadrio_ricco_2004}, plasma actuators \citep{Choi2011}, and synthetic jets \citep{Glezer2002}. Due to the complexity of reducing drag, it has been most common to apply these control methods in an open-loop manner where the control policy at any given time is independent of the flow state \cite{Choi2008}. However, application of feedback control on a turbulent system could yield far better performance in controlling drag. 

Unfortunately, the complexity of the problem has typically limited applications of feedback control to methods based on heuristics. A well-studied heuristic method is opposition control \citep{Choi1994}. Here the wall-normal velocity at the wall is set to have the opposite sign as the wall-normal velocity at some detection plane in the channel, a straightforward actuation in simulations. This method has been applied in simulations \citep{Choi1994,Hammond1998,Chung2011,Ibrahim2019} and experiments (with some modifications) \citep{Rebbeck2006,Cheng2021}, and extensions exist to use just wall observations \citep{Lee1997,Park2020}. When heuristics are replaced with methods from optimal control theory, like model predictive control (MPC),  the drag reduction far outperforms opposition control \cite{Bewley2001} \KZrevise{while using the same actuation scheme.} However, the real-time implementation of MPC on DNS \ALrevise{still remains infeasible because it involves solving the DNS forward over a time horizon (preferably a long one) and then solving an adjoint backwards in time for every actuation \citep{Bewley2001}.}

A potential approach to overcome the high computational cost of real-time optimization of a control strategy is deep reinforcement learning (RL) \citep{SuttonBarto2018}. Deep RL gained significant traction when it was used to defeat the best professional players in GO \citep{Silver2016a}, DOTA II \citep{OpenAI2019}, and Starcraft II \citep{Vinyals2019}, in addition to the best engines in GO, Chess, and Shogi \citep{Silver2018}. In deep RL, a neural network (NN) control agent is trained through iterative interactions with the environment (i.e.~the system to be controlled) to maximize a scalar total reward (i.e.~control objective) that includes present as well as discounted future reward values. Once trained, the control agent can be deployed in real time without the need for online optimization. 

In recent years, RL has been applied in fluids simulations to reduce the drag experienced in flow around a cylinder \citep{Rabault2019,Li2022,Varela2022}, to optimize jets on an airfoil \citep{Wang2022}, and to find efficient swimming strategies \citep{Verma2018}. RL has even been applied to experimental flow systems \cite{Fan2020}. 
Recently multi-agent deep RL has been explored for the control of pressure-driven turbulent channel flow \citep{Sonoda2022,Guastoni2023} in a problem formulation similar to opposition control \citep{Choi1994}. In these works, an RL policy is trained to map local detection plane observables to a wall-normal velocity response at the walls to reduce drag. Notably, the same RL policy is locally implemented at each wall grid point. We differentiate the control problem addressed from the previously mentioned works in that we limit the control authority to just two spatially localized jets on a single wall, with a zero-net flux constraint, as opposed to full spatial control of both walls. We feel that this is much closer to experimental realizability than an approach with control authority everywhere on the wall. A recent review of the application of deep RL applied to fluid mechanics problems is presented in \citet{Viquerat2023}. 

In these active flow control problems, deep RL possesses the advantageous property of being completely data-driven, allowing it to discover novel and nontrivial control strategies in complex systems from just data alone without the need to analytically derive or hard-code system properties into the method. 
However, the training portion of RL is a major bottleneck, requiring a tremendous number of interactions with the target environment to find an approximately optimal policy \citep{Dulac-Arnold2021}. Practically speaking, this can correspond to running an enormous number of high-resolution simulations or flow experiments, both of which may be prohibitively expensive.


In the present work, we apply RL to control a minimal flow unit (MFU) (the smallest domain that sustains turbulence) \citep{Jimenez1991} of plane Couette flow at $Re=400$ using a pair of streamwise-aligned slot jets at one wall, with a no-net-flux constraint. Therefore there is only one degree of freedom for actuation. We select this system because the unactuated flow isolates the self-sustaining regeneration cycle of wall-bounded turbulence \citep{Hamilton1995,Inubushi2015}. This case is well-studied for tasks such as reduced-order modeling \cite{Waleffe1997a,Moehlis2004,Gibson2002,Linot2023}, finding invariant solutions \citep{Gibson2008,Viswanath2007}, and applying opposition control  \cite{Ibrahim2019}.


In order to overcome the high computational cost of RL training in this environment, in this work we replace the high-resolution simulation with an accurate low-dimensional surrogate model, aiming to dramatically reduce the time required to train the control policy.
We showed in \cite{ZengRSPA2022} that this data-driven model-based RL approach, which we refer to as ``Data-Driven Manifold Dynamics" RL (DManD-RL), works well for controlling spatiotemporal chaotic dynamics in the Kuramoto-Sivashinksy Equation. For further discussion on the various types of model-based RL, we refer the reader to \citet{ZengRSPA2022}. In Sec.\ \ref{sec:Framework} we introduce the control environment and the DManD-RL framework. Then, in Sec.\ \ref{sec:Results} we describe the data used for training the DManD model, the performance of the model, and the results of applying RL to the DManD model and to the DNS environment. Finally, we conclude in Sec.\ \ref{sec:Conclusions} with a summary of the key results.

\section{Framework} \label{sec:Framework}

\subsection{Navier-Stokes Equation with Slot Jets}

\begin{figure} 
    \centering
	\includegraphics[trim=0 0 0 0,width=\textwidth,clip]{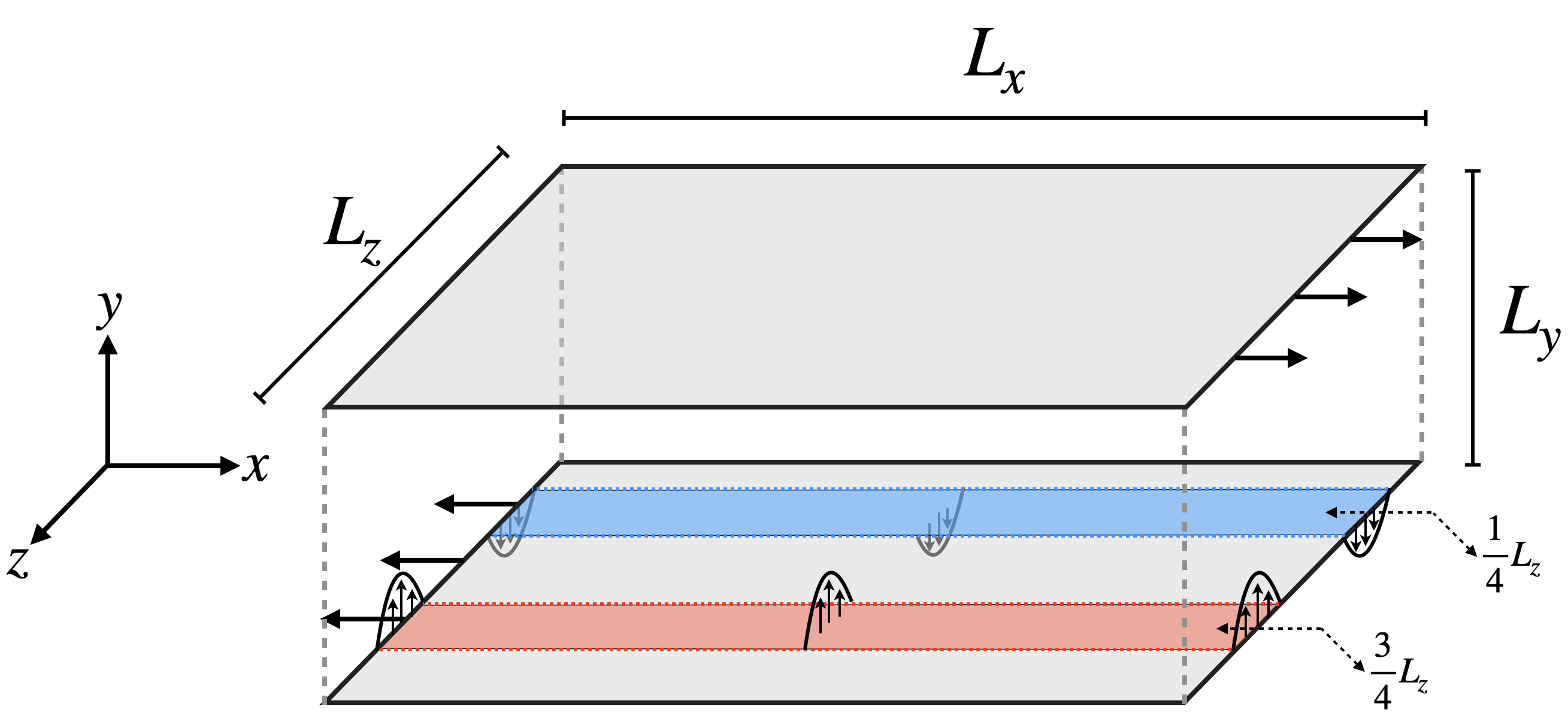}
	\captionsetup[subfigure]{labelformat=empty}
	\caption{Schematic of the Couette flow domain with two slot jets on one wall. }
	\label{fig:CouetteSchematic}
\end{figure} 
The environment we consider is a direct numerical simulation (DNS) of the Navier-Stokes Equations (NSE)
\begin{equation} \label{eq:NSE}
	\dfrac{\partial \mathbf{u}}{\partial t}+\mathbf{u}\cdot \nabla \mathbf{u}=-\nabla p+\Rey^{-1}\nabla^2 \mathbf{u}, \quad \nabla \cdot \mathbf{u}=0.
\end{equation} 
\ALrevise{The }velocities in the streamwise $x\in[0,L_x]$, wall-normal $y\in[-L_y/2,L_y/2]$, and spanwise $z\in[0,L_z]$ directions \ALrevise{are} defined as $\mathbf{u}=[u_x,u_y,u_z]$, and \ALrevise{the pressure is} $p$. \ALrevise{Here we have nondimensionalized velocity by the speed $U$ of the walls, length by the channel half-height ($h=L_y/2$), time with $h/U$ and pressure with $\rho U^2$, where $\rho$ is the fluid density. The Reynolds number is $\Rey=Uh/\nu$, where $\nu$ is the kinematic viscosity.} 
 The boundary conditions for this setup are periodic in $x$ and $z$ ($\mathbf{u}(0,y, z)=\mathbf{u}(L_x,y, z), \mathbf{u}(x,y, 0)=\mathbf{u}(x,y, L_z)$), no-slip boundary conditions at the walls ($u_x(x,\pm L_y/2, z)=\pm 1, u_z(x,\pm L_y/2, z)=0$), no penetration at the top wall ($u_y(x,L_y/2,z)=0$), and finally, the actuation on the bottom wall ($u_y(x,-L_y/2, z)=f_a(x,z)$), as we now describe. 

The actuation on the bottom wall is in the form of two slot jets that are Gaussian in $z$ and travel the length of the channel:
\begin{equation} \label{eq:jets}
    u_y(x,-L_y/2, z)=f_a(x,z)=a(t)V_\text{max}\left(\text{exp}\left(-\dfrac{(z-L_z/4)^2}{2\sigma^2}\right)-\text{exp}\left(-\dfrac{(z-3L_z/4)^2}{2\sigma^2}\right)\right).
\end{equation}
We set $\sigma\approx 0.16$ so that the jets act ``locally", and the velocity of the jet is dictated by $a(t) V_\text{max}$, where $a(t)\in[-1,1]$ is the instantaneous actuation amplitude scaled by a maximum velocity $V_\text{max}=0.05$. \ALrevise{For perspective, the root-mean-squared wall-normal velocity at the channel centerline for turbulent unactuated flow is $\sim0.063$.}
 We chose this small velocity to evaluate how the agent performs with limited control authority. In Fig.\ \ref{fig:CouetteSchematic} we show a schematic illustrating this system.

The complexity of the flow increases as the Reynolds number 
and the domain size $L_x$ and $L_z$ increase. Here we chose the same setup as \citet{Hamilton1995}, $\Rey=400$ and $[L_x,L_y,L_z]=[1.75\pi,2,1.2\pi]$. These parameters isolate the ``self-sustaining process" (SSP) that drives wall-bounded turbulence. In the SSP, low-speed streaks that have been lifted from the wall become wavy, this waviness leads to the breakdown of the streaks, generating streamwise rolls, and, finally, these rolls lift low-speed fluid off the wall to regenerate streaks, completing the cycle. By working in this well-studied domain that is dominated by the SSP, we can better identify the means by which a control strategy can disrupt or suppress this process.

In this work, the control strategy is to minimize the turbulent drag averaged between both walls
\begin{equation}
    D=\dfrac{1}{2}\int_0^{L_x}\int_0^{L_z} \left(\left.\dfrac{\partial u_x}{\partial y}\right|_{y=1}-1\right) +\left(\left.\dfrac{\partial u_x}{\partial y}\right|_{y=-1}-1\right)dxdz,
\end{equation}
\MDGrevise{subject to a quadratic penalty on actuation amplitude $a$. (If the relation between the pressure drop and actuation velocity for pumping fluid into/out of the domain is linear, then this penalty is proportional to the power consumption of the actuation.) Further details are described in Sec.~\ref{sec:RL1}.}
We report drag in this fashion because this quantity goes to 0 when the flow laminarizes. 


We simulate the flow using a Fourier-Chebyshev pseudo-spectral code we implemented in Python \citep{Linot2023_git}, which is based on the \emph{Channelflow} code developed by Gibson et al. (\citep{Gibson2012,Gibson2021}). 
Linear terms are treated implicitly and the nonlinear term explicitly.  The specific time integration schemes we use are the multistage SMRK2 scheme \citep{Spalart1991} for the first two timesteps after every actuation, and the multistep Adams-Bashforth Backward-Differentiation 3 scheme \citep{Peyret2002} until the next actuation. The multistep scheme is more computationally efficient, but, because actuations change instantaneously, using previous steps with the incorrect boundary condition would lead to incorrect results. For all trials we evolve solutions forward using $\Delta t=0.02$ on a grid of $[N_x,N_y,N_z]=[32,35,32]$ in $x$, $y$, and $z$ from random divergence-free initial conditions that we evolve forward $100$ time units so initial conditions are near the turbulent attractor.

While most of this approach is standard, here we include some details on the simulation procedure to highlight explicitly how we set the jet actuation boundary condition. At each time step, the approach involves solving the expression
\begin{equation} \label{eq:NSEtimestep}
	\Rey^{-1} \dfrac{d^2\hat{\mathbf{u}}_{k_x,k_z}^{i+1}}{dy^2}-\lambda \hat{\mathbf{u}}_{k_x,k_z}^{i+1} -\hat{\nabla}\hat{p}_{k_x,k_z}^{i+1}=-\hat{\mathbf{R}}_{k_x,k_z}^{i}, 
\end{equation} 
where $i$ is the timestep and $\hat{\cdot}=\mathcal{F}_{x,z}(\cdot)$ denotes the Fourier transform in $x$ and $z$. The variable $\lambda$ includes the timestep $\Delta t$ and the $x$ and $z$ components of the diffusive term and $\mathbf{R}$ encompasses all the remaining explicit terms (for a multistep method this includes $\hat{\mathbf{u}}_{k_x,k_z}$ multiple steps back). We refer the reader to \cite{Gibson2012} for a more detailed discussion. Upon taking the divergence of Eq.\ \ref{eq:NSEtimestep}, and accounting for incompressibility, we isolate the problem down to 4 sets of one-dimensional Helmholtz equations (for conciseness we suppress indices $k_x$, $k_z$, and $i$):

\begin{align}
    \Rey^{-1} \dfrac{d^2\hat{u}_x}{dy^2}-\lambda \hat{u}_x -\dfrac{2\pi i k_x}{L_x}\hat{p}=-\hat{R}_x& &\hat{u}_x(\pm1)= \pm \delta_{k_x,0}\delta_{k_z,0} \label{eq:u}\\
    \Rey^{-1} \dfrac{d^2\hat{u}_z}{dy^2}-\lambda \hat{u}_z -\dfrac{2\pi i k_z}{L_z}\hat{p}=-\hat{R}_z& & \hat{u}_z(\pm1)=0 \label{eq:w}\\
    \Rey^{-1} \dfrac{d^2\hat{u}_y}{dy^2}-\lambda \hat{u}_y -\dfrac{d\hat{p}}{dy}=-\hat{R}_y& & \hat{u}_y(-1)=\mathcal{F}_{x,z}(f_a), \quad \hat{u}_y(1)=0 \label{eq:v}\\
    \dfrac{d^2\hat{p}}{dy^2}-4\pi^2\left(\dfrac{k_x^2}{L_x^2}+\dfrac{k_z^2}{L_z^2} \right)\hat{p}=\hat{\nabla} \cdot \hat{\mathbf{R}}& &\dfrac{d\hat{u}_y}{dy}(\pm1)=0. \label{eq:p}
\end{align}


These equations can be solved for every wavenumber pair $k_x$ and $k_z$. The challenge in solving these equations is due to the coupling in Eq.\ \ref{eq:v} and Eq. \ref{eq:p}. The pressure is coupled to the wall-normal velocity because an explicit boundary condition is unknown.
Instead, from incompressibility, we know $d\hat{u}_y/dy(\pm1)=0$, which we substitute for the pressure boundary condition. To solve these coupled equations we use the influence matrix method and tau correction developed by \citet{Kleiser1980}. Although we set the wall-normal boundary condition in Eq. \ref{eq:v} by the slot jets in Eq.\ \ref{eq:jets}, we note that it is simple to replace this boundary condition with any shape of actuation.

\subsection{Data-driven framework} \label{sec:RL1}
The objective in deep RL is to train an agent, commonly a neural network, to approximate the optimal control policy $a=\pi^*(s)$, which given a state observation $s$, outputs the optimal control action $a$. The optimal policy seeks to maximize the expected long time discounted cumulative reward,
\begin{equation}
	\pi^*=\arg\max_\pi \mathbb{E}\left[\sum_{l=0}^{\infty} \gamma^l (r_{t+l\tau}) \right],
 \label{eq:cumulativereward}
\end{equation}
where $0<\gamma<1$ is the discount factor, $\tau$ is the time between control actions, and $r_t$ is the reward, a scalar-valued control objective function decided by the user evaluated at time $t$. As our objective in this work is to minimize the drag of our turbulent Couette system while simultaneously avoiding the use of superfluous control actions, we define the reward function as the following,
\begin{equation}
	r_t=- \left<D(t) + c \|a(t)\|^2 \right>_{\tau}, 
 \label{eq:reward}
\end{equation}
where $c$ is a scalar and $\langle\cdot\rangle_{\tau}$ is the average from $t$ to $t+\tau$. \KZrevise{We note here that our actuation penalty is proportional to the power required for actuation}.
In deep RL $\pi^*$ is learned via repeated cyclic interactions between the agent and the environment i.e.\ the target system. A typical cycle consists of the following: given a state observation of the system at time $t$, $s_t$, the control agent outputs its estimated best control response $a_t$. This control action is then applied to the environment. The system is allowed to evolve for $\tau$ time units, and then the impact of the action is quantified by observing the resulting system state, $s_{t+\tau}$, as well as the reward signal, $r_t$. This iterate of data, [$s_t$,$a_t$,$r_t$,$s_{t+\tau}$], is then stored and used for updating the control agent for the next time interval.


\subsection{DManD Modeling Framework} \label{sec:FrameworkModeling}

Applications of deep RL often require repeating this cycle $\mathcal{O}(10^{6+})$ times. Because deep RL conventionally requires an online realization of the target system during training, the practicality of training an RL agent for systems that are computationally or experimentally expensive to realize online, e.g.~a DNS of turbulent channel flow, is especially bottlenecked by the expense of the environment itself \citep{Dulac-Arnold2021}.

\begin{figure} 
    \centering
	\includegraphics[trim=0 0 0 0,width=\textwidth,clip]{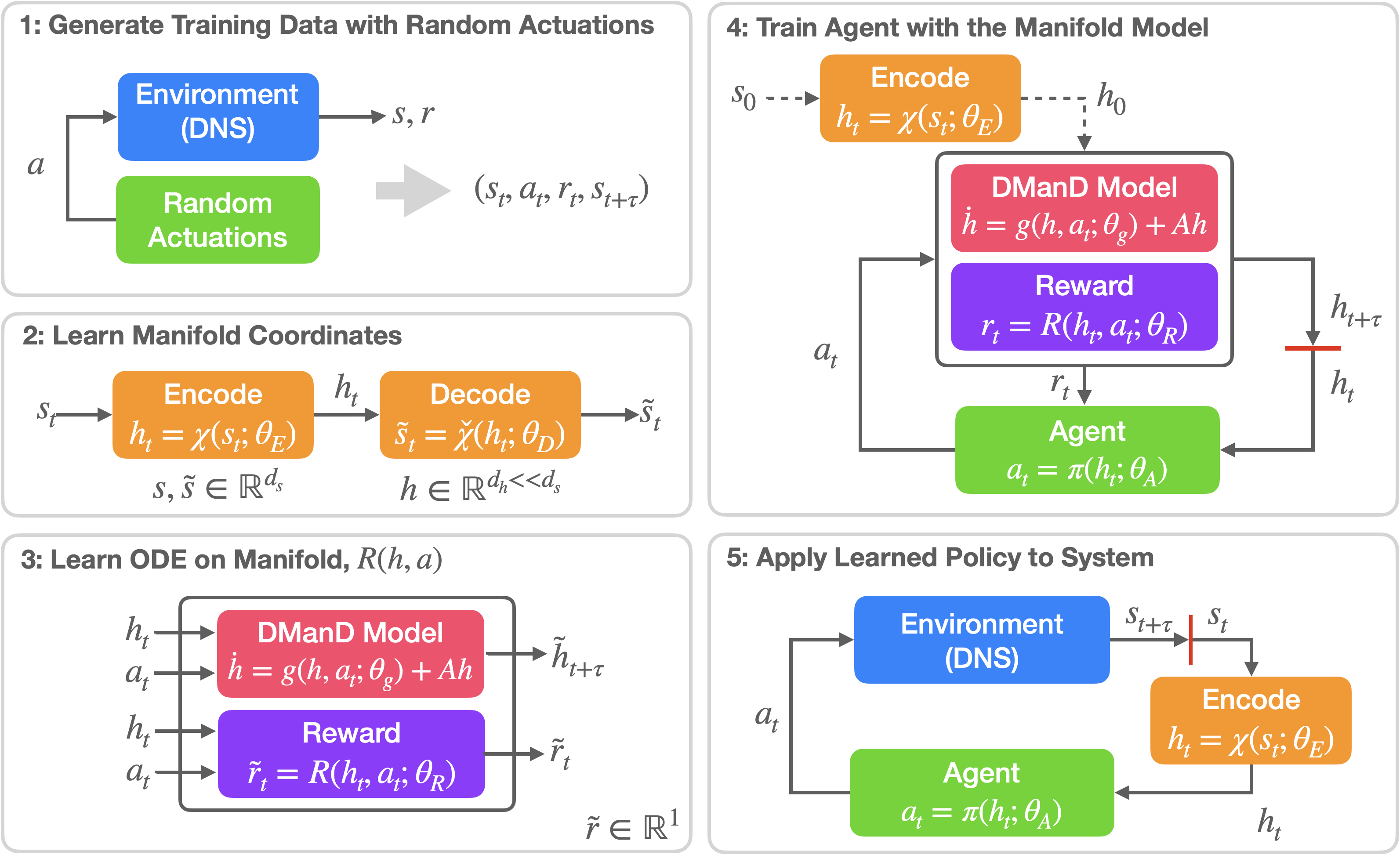}
	\captionsetup[subfigure]{labelformat=empty}
	\caption{Schematic of the DManD-RL framework. After step 3, $\tilde{\cdot}$ is omitted for clarity.}
	\label{fig:MethodSchematic}
\end{figure} 

To circumvent this bottleneck, we employ a method denoted ``Data-driven Manifold Dynamics for RL" \citep{ZengRSPA2022}, or ``DManD-RL" for short, with some modification. This framework consists of two main learning objectives, which can be broken down into five steps, illustrated in Fig.\ \ref{fig:MethodSchematic}. The first objective is to obtain an efficient and accurate low-dimensional surrogate model of the underlying dynamics of the turbulent DNS, which we refer to as the DManD model. This objective is achieved via the first three steps outlined in Fig.\ \ref{fig:MethodSchematic}: 1) collect data tuples of the target system experiencing random control actions, 2) obtain a low-dimensional representation of the environment's dynamics, 3) model the dynamics of the environment and its response to control inputs.

The second objective is to use this DManD model for RL training to quickly and efficiently obtain an effective control agent. This objective is achieved via the remaining two steps outlined in Fig.\ \ref{fig:MethodSchematic}: 4) perform deep RL with the DManD model, and 5) deploy the control agent to the original environment. In the following sections we discuss the details for generating the DManD model in Sec.\ \ref{sec:FrameworkModeling} and the method for training and deploying the DManD-RL agent in Sec.\ \ref{sec:FrameworkRL}. 

As this framework is completely data-driven, the first step involves collecting sufficient data to learn an accurate surrogate model. This model must capture the underlying flow system, the response of the dynamics to control inputs, and the impact the control inputs have on the objective. Generating this model requires that we have a large dataset that includes the cycle of data described above ([$s_t$,$a_t$,$r_t$,$s_{t+\tau}$]). In RL training actions are chosen by the policy, however, for training the model we do not necessarily have any policy to generate this data. As such, we instead chose to randomly actuate the flow to generate the original dataset used in training the DManD model. Details on the specifics of the dataset are included in Sec.\ \ref{sec:Resultsa}.  

With this data, the second step of the DManD-RL framework involves finding a low-dimensional representation of the state. For many dissipative systems, there is either proof or evidence that the long-time dynamics collapse onto a finite-dimensional invariant manifold\cite{Foias1988a,Doering1988,Zelik2014,Foias1988,Temam1989}. 
We can define a mapping to coordinates parameterizing this manifold
\begin{equation}
	h_t=\chi(s_t),
\end{equation} 
where $h_t\in\mathbb{R}^{d_h}$ is the manifold coordinate system and an inverse mapping back to the state
\begin{equation}
	s_t=\check{\chi}(h_t).
\end{equation}

When the data lies on a finite-dimensional invariant manifold then the finite-dimensional manifold coordinate representation $h_t$ contains the same information as the state $s_t$. Thus, if we know $\chi$ and $\check{\chi}$ we can simply use $h_t$ in place of $s_t$ for training the RL agent, which requires far fewer degrees of freedom. 
\ALrevise{One subtlety that we gloss over here is that a $d_\mathcal{M}$-dimensional manifold may require a set of overlapping local representations called \emph{charts} if one wants to represent the manifold with $d_\mathcal{M}$ parameters \cite{Lee2003,floryan2021charts,Fox2023}.
However, a manifold with a topological dimension $d_\mathcal{M}$ can be embedded in $\mathbb{R}^{2d_\mathcal{M}}$ \citep{Whitney1944,Sauer1991}. So, in the worst case, as long as $d_h\geq 2d_\mathcal{M}$, a single global coordinate representation can be used, as we do here.}

In this work we will approximate $\chi$ and $\check{\chi}$ using an undercomplete autoencoder. 
This consists of two NNs: an encoder ($\chi$) that reduces the dimension and a decoder ($\check{\chi}$) that expands it. Here we train an autoencoder to find the correction from the linear map given by the proper orthogonal decomposition (POD) \cite{Linot2020,Linot2021}. As such we define the state observation $s$ to come from projecting the flow field $\mathbf{u}$ onto a set of POD modes \ALrevise{(i.e.\ $s$ is the POD coefficients)}. Sec.\ \ref{sec:Resultsa} includes details on our POD implementation.

\ALrevise{For the encoding, we sum the leading $d_h$ POD coefficients with a correction from a NN that is a function of all the POD coefficients:}
\begin{equation}
	h_t=\chi(s_t;\theta_E)=s_{t,d_h}+\mathcal{E}(s_t;\theta_E),\label{eq:encode}
\end{equation}
where $s_{t,d_h}$ is the first $d_h$ components of $s_t$ and $\mathcal{E}$ is a NN. 
\ALrevise{For the decoding, we want to reconstruct all $500$ POD coefficients from the $d_h$ values we have from the encoding. These $d_h$ values are approximately the leading POD coefficients so we can again just add these values to a NN that corrects the leading $d_h$ POD coefficients and reconstructs the remaining POD coefficients:}
\begin{equation}
	\tilde{s}_t=\check{\chi}(h_t;\theta_D)=[h_t,0]^T+\mathcal{D}(h_t;\theta_D).\label{eq:decode}
\end{equation}
Here, $[h_t,0]^T$ represents $h_t$ padded with zeros to the correct size, and $\mathcal{D}$ is a NN. The notation $\tilde{\cdot}$ indicates that this is an approximation of $s_t$.
We refer to this autoencoder structure as a hybrid autoencoder, in contrast to the standard approach of simply treating $\chi(s_t;\theta_E)$ and $\check{\chi}(h_t;\theta_D)$ as NNs. We take this approach because it can achieve lower reconstruction errors \cite{Linot2020} than the standard approach and the variables $h$, and is more interpretable, as a nonlinear correction to POD.

The NNs $\mathcal{E}$ and $\mathcal{D}$ are trained to minimize 
\begin{equation} \label{eq:LossAuto}
	L=\dfrac{1}{dK}\sum_{i=1}^K||s_{t_i}-\check{\chi}(\chi(s_{t_i};\theta_E);\theta_D)||_2^2 +\dfrac{1}{d_hK}\sum_{i=1}^K\xi ||\mathcal{E}(s_{t_i};\theta_E)+\mathcal{D}_{d_h}(h_{t_i};\theta_D)||_2^2,
\end{equation}
where $\mathcal{D}_{d_h}$ is the first $d_h$ components of the decoder, $\xi$ is a scalar, and \ALrevise{$K$ is the batch size}. In this loss, the first term is the reconstruction mean-squared error (MSE), and the second term promotes the accurate reconstruction of the leading $d_h$ POD coefficients. \ALrevise{We include this second term because it must go to zero if the reconstruction is perfect. This is because the modification that the NN makes in the encoder must be removed by the NN in the decoder.}
In Sec.\ \ref{sec:Resultsa2} we provide details on autoencoder training. 

Now that we have a low-dimensional representation of the state observation, in step three, we train a model to predict the evolution of $h$ and the reward $r_t$. To predict the evolution of $h$ (from $h_t$ to $h_{t+\tau}$) we train a ``stabilized" neural ordinary differential equation (ODE) \citep{Linot2022}
\begin{equation}\label{eq:ODENet_Damp}
	\dfrac{dh}{dt}=g(h,a;\theta_g)+A h,
\end{equation}
which can be integrated forward in time to predict
\begin{equation}\label{eq:ODENet_Int}
	\tilde{h}_{t+\tau}=h_t+\int_{t}^{t+\tau}g(h,a;\theta_g)+A h dt,
\end{equation}
where $g$ is a NN and $A$ is a matrix that can be learned from data or fixed. We chose to find an ODE instead of a discrete timestepping method because it allows us to evolve this equation to arbitrary times that may not align with the sampling rate of our data. This is a highly desirable trait as it means we can freely vary the time between actions after training the DManD model, which could not be done if we found a discrete time map from $h_t$ to $h_{t+\tau}$. 

The linear term in Eq.\ \ref{eq:ODENet_Damp} is important for stability. Without this term, small errors in the dynamics can lead to linear growth at long times, which a linear damping term prevents \citep{Linot2021,Linot2022,Linot2023}. In this work, we set the linear term 
\begin{equation} \label{eq:Damp}
	A_{ij}=-\beta \delta_{ij}\sigma_i(h),
\end{equation} 
where $\beta=0.1$, $\delta_{ij}$ is the Kronecker delta, and $\sigma_i(h)$ is the standard deviation of the $i$th component of $h$. This term acts as a damping,  preventing trajectories from moving far away from the training data. In \cite{Linot2023} we show in Couette flow without actuation that this term prevents models from becoming unstable. \ALrevise{We note that the addition of this damping term does not negatively impact the accuracy of the vector field in the region of state space where there is data, because the damping term is present when we train the NN. This allows the NN to compensate for the damping term in the region where the data lies.}  We train the NN $g$ to minimize  
\begin{equation} \label{eq:LossODE}
	J=\dfrac{1}{d_hK}\sum_{i=1}^K ||h_{t_i+\tau}-\tilde{h}_{t_i+\tau}||_2^2.
\end{equation}
We describe in more detail how the gradient of this loss is computed in \cite{Linot2023}.


The final piece of the algorithm is computing the reward. We could to this directly by mapping $h$ back to $\mathbf{u}$ and computing the corresponding drag. However, this is undesirable because it is computationally expensive, and, in general, it may not be possible to directly compute the reward from the state. To overcome this difficulty, we use a NN to compute an estimate $\tilde{D}$ of the drag from $h$:
\begin{equation} 
	\tilde{D}=\mathcal{R}(h;\theta_R),\label{eq:rewardnetwork}
\end{equation}
which we train to minimize $J_D=1/K \sum_{i=1}^K||D_{t_i}-\tilde{D}_{t_i}||_2^2$. Then, we compute the reward $\tilde{r}_t=R(h_t,a_t;\theta_R)$ by inserting $\tilde{D}$ into Eq.\ \ref{eq:reward}.

\subsection{Reinforcement Learning using DManD} \label{sec:FrameworkRL} 

With our DManD model of the underlying dynamics, we can now proceed to step four and efficiently obtain a control policy by training an RL agent to interact with the low-dimensional DManD model rather than the expensive DNS. Rather than learning $a_t=\pi(s_t;\theta_A)$, our goal shifts to learning
\begin{equation} \label{eq:policyh}
	a_t=\pi(h_t;\theta_A),
\end{equation}
where $\theta_A$ are the network parameters of the RL agent. In this work, we employ the Soft Actor-Critic (SAC) RL algorithm \citep{SAC} but we emphasize that the DManD-RL framework works with any general RL algorithm. SAC was chosen in this application because it possesses several advantageous characteristics including the ability to output control signals from a continuous range, an off-policy formulation that allows the ``reuse" of previously generated data, and twin critic networks to aid the brittleness commonly associated with many off-policy deep RL algorithms. Distinctively, SAC has a stochastic actor with an additional entropy-maximizing formulation. This formulation modifies the typical RL objective of Eq. \ref{eq:cumulativereward} to the following
\begin{equation}
	\pi^*=\arg\max_\pi \mathbb{E}\left[\sum_{l=0}^{\infty} \gamma^l \left( r_{t+l\tau} + \alpha\mathcal{H}(\pi(\cdot|s_t)) \right) \right],
 \label{eq:cumulativerewardSAC}
\end{equation}
where the entropy of the policy, $\mathcal{H}$, is defined as
\begin{equation}
	\mathcal{H}(\pi(\cdot|s_t))=-\log(\pi(\cdot|s_t)).
 \label{eq:entropyH}
\end{equation}
 Here $\alpha$ is the trade-off coefficient, which is set to 1.0. This entropy-regularized objective promotes wider state-action exploration and the ability to invest in multiple modes of near-optimal strategies in addition to maximizing the cumulative reward. In this work, the agent is trained stochastically but deployed deterministically during testing. Our implementation of SAC utilizes NNs to approximate the policy function (i.e.\ agent) $\pi$, the two critic functions, $Q_1$ and $Q_2$, and the value function $V$. For more details regarding the derivation and implementation of SAC, we refer the reader to \citet{SAC}.

With a trained DManD-RL agent, we can proceed to the fifth and final step and deploy the control agent to the original turbulent channel DNS for application. As the RL agent learned its control strategy by observing the low dimensional manifold coordinate system, we must map the high-dimensional state observations made in the DNS to the proper input by using the previously obtained encoder function, $\mathcal{\chi}$,
\begin{equation}
	a_t=\pi(\chi(s_t)).
 \label{eq:deployment}
\end{equation}
The trained agent can now be deployed in a closed-loop control fashion.

\section{Results} \label{sec:Results}
\subsection{Description of Data} \label{sec:Resultsa}

The first step in DManD-RL is to generate a data set for training models. Our data set consists of $500$ different initial conditions, which we evolved forward $300$ time units with a random actuation chosen every $5$ time units. We sample the random actuations uniformly between $-1$ and $1$. Every $1$ time unit we record the velocity field $\mathbf{u}$ and the action $a$, yielding $1.5\cdot 10^5$ snapshots of data. We use $80\%$ of this data to train the models, and the remaining $20\%$ to test performance on the data never previously seen by the model.

Due to the high-dimensional nature of the data, learning a manifold coordinate system using the velocity field on the grid as the state is challenging. As such, we first preprocess the data using the proper orthogonal decomposition (POD) to reduce the dimension from $\mathcal{O}(10^5)$ to $500$ and treat these $500$ POD coefficients as the state observation $s$. 
In the POD, we find modes $\boldsymbol{\Phi}$, which we project onto to maximize
\begin{equation}
	\dfrac{\left<\left|(\mathbf{u}',\boldsymbol{\Phi})\right|^2\right>}{||\boldsymbol{\Phi}||^2},
\end{equation} 
which uses the fluctuating velocity $\mathbf{u}'=\mathbf{u}-\left< \mathbf{u}\right>$ ($\left<\cdot\right>$ denotes the average).
As shown in \cite{Smith2005,Holmes1998}, these modes can be found by solving the following eigenvalue problem:
\begin{equation} \label{eq:Direct}
	\sum_{j=1}^3 \int_0^{L_x}\int_{-1}^1 \int_0^{L_z}
	\left\langle u'_i(\mathbf{x}, t) \bar{u}_j\left(\mathbf{x}^{\prime}, t\right)\right\rangle \Phi_j^{(n)}\left(\mathbf{x}^{\prime}\right) d\mathbf{x}^{\prime}=\lambda_i \Phi_i^{(n)} (\mathbf{x}),
\end{equation} 
where $\bar{\cdot}$ is the complex conjugate. Na\"ive implementation of POD requires solution of a computationally expensive $d \times d$ eigenvalue problem after approximating these integrals. This formulation also fails to respect the streamwise translation invariance of our system. 

We account for this translational invariance, and make the problem tractable, by exploiting the fact that in translation invariant directions POD eigenfunctions take the form of Fourier modes \cite{Holmes1998}. Note that the fixed position of the slot jets breaks translation invariance in $z$. 
This turns the eigenvalue problem in Eq.\ \ref{eq:Direct} into
\begin{equation}  \label{eq:Direct2}
L_x \sum_{j=1}^3 \int_{-1}^1 \int_0^{L_z}\left\langle \hat{u}'_i(k_x,y^{\prime},z^{\prime},t) \bar{\hat{u}}_j^{\prime}(k_x,y^{\prime},z^{\prime},t)\right\rangle \varphi_{jk_x}^{(n)}\left(y^{\prime},z^{\prime}\right) d y^{\prime}d z^{\prime}=\lambda_{k_x}^{(n)} \varphi_{ik_x}^{(n)}(y,z).
\end{equation}
This is a $3 N_yN_z \times 3 N_yN_z$ eigenvalue problem for every wavenumber $k_x$. We speed up the computation of this eigenvalue problem by evenly sampling $5,000$ snapshots of training data.

\begin{figure} 
	\includegraphics[trim=0 0 0 0,width=\textwidth,clip]{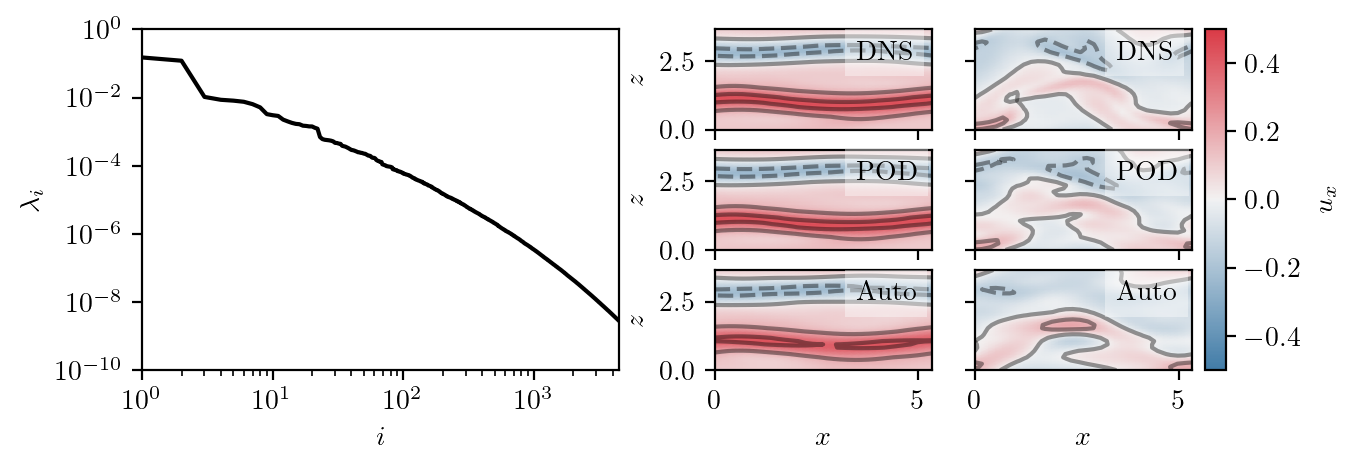}
	\captionsetup[subfigure]{labelformat=empty}
	\begin{picture}(0,0)
	\put(0,162){\contour{white}{ \textcolor{black}{a)}}}
	\put(214,162){\contour{white}{ \textcolor{black}{b)}}}
 	\put(321,162){\contour{white}{ \textcolor{black}{c)}}}
	\put(214,121){\contour{white}{ \textcolor{black}{d)}}}
 	\put(321,121){\contour{white}{ \textcolor{black}{e)}}}
	\put(214,80){\contour{white}{ \textcolor{black}{f)}}}
    \put(321,80){\contour{white}{ \textcolor{black}{g)}}}
	\end{picture} 
	\begin{subfigure}[b]{0\textwidth}\caption{}\vspace{-10mm}\label{fig:SVExa}\end{subfigure}
	\begin{subfigure}[b]{0\textwidth}\caption{}\vspace{-10mm}\label{fig:SVExb}\end{subfigure}
	\begin{subfigure}[b]{0\textwidth}\caption{}\vspace{-10mm}\label{fig:SVExc}\end{subfigure}
	\begin{subfigure}[b]{0\textwidth}\caption{}\vspace{-10mm}\label{fig:SVExd}\end{subfigure}
	\begin{subfigure}[b]{0\textwidth}\caption{}\vspace{-10mm}\label{fig:SVExe}\end{subfigure}
	\begin{subfigure}[b]{0\textwidth}\caption{}\vspace{-10mm}\label{fig:SVExf}\end{subfigure}
 	\begin{subfigure}[b]{0\textwidth}\caption{}\vspace{-10mm}\label{fig:SVExg}\end{subfigure}
  
	\caption{(a) Eigenvalues from the POD. (b) and (c) Snapshots of the centerline streamwise velocity ($u_x(y=0)$) from the DNS at representative low-drag (left) and high-drag (right) instants. (d) and (e) POD reconstruction with 500 modes of the DNS results in (b) and (c). (f) and (g) Autoencoder reconstruction of (d) and (e). Solid contour lines are positive and dotted contour lines are negative.}
	\label{fig:SVEx}
\end{figure}
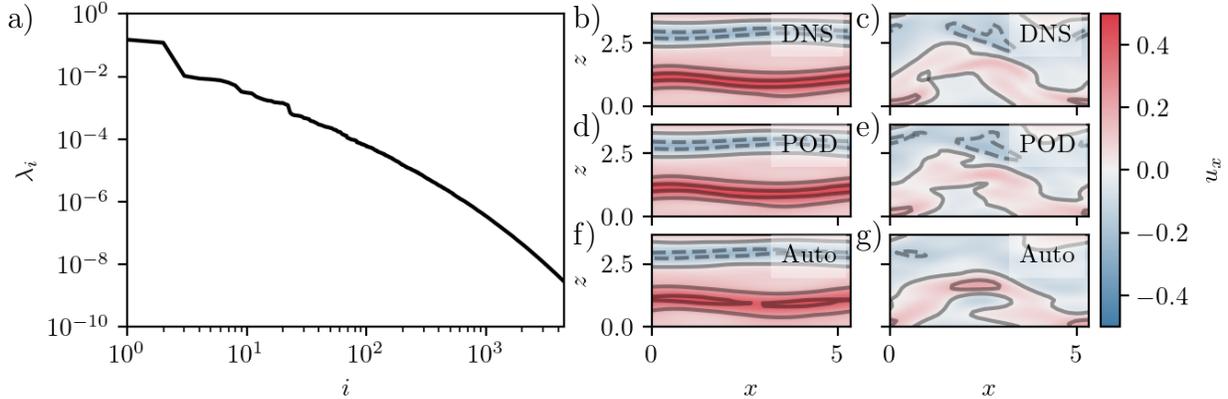

Solving the eigenvalue problem in Eq.\ \ref{eq:Direct2} results in eigenvectors
\begin{equation}
	\boldsymbol{\Phi}_{k_x}^{(n)}(\mathbf{x})=\frac{1}{\sqrt{L_x}} \exp \left(2 \pi i\frac{k_x x}{L_x}\right) \boldsymbol{\varphi}_{k_x}^{(n)}(y,z),
\end{equation}
and eigenvalues $\lambda_{k_x}^{(n)}$. We then project $\mathbf{u}$ onto the leading $305$  modes, sorted according to the magnitude of $\lambda$, giving us the state observation $s$. This results in a $500$-dimensional state because a majority of the modes are complex. Figure \ref{fig:SVExa} shows the sorted eigenvalues. The eigenvalues drop off quickly resulting in the first $305$ modes containing $99.8\%$ of the energy.  

\subsection{Manifold Coordinate System} \label{sec:Resultsa2}

Now that we reduced the dimension of the state observation $s$, we train an autoencoder as described in Sec.\ \ref{sec:FrameworkModeling} to find $h$. In \citet{Linot2023} we varied $d_h$ for the unactuated Couette flow system and found the DManD models to be highly accurate with fewer than $20$ degrees of freedom. In the present system, additional degrees of freedom will be excited by the actuations, so we increased $d_h$ slightly, to $d_h=25$. For training, we first normalized $s$ by subtracting the mean and dividing by the maximum standard deviation. \KZrevise{We then trained four autoencoders until the training error stopped improving. We selected the autoencoder based on which yielded the best forecasting performance when coupled to a neural ODE described in the following section. The selected autoencoder had a test MSE of $1.45\cdot 10^{-4}$.} The architecture and parameters of the autoencoders are reported in Table \ref{Table1}. 

In Figs.\ \ref{fig:SVExb}-\ref{fig:SVExg} we compare the centerplane streamwise velocity of two flowfields -- one in a bursting state (high-drag) and the other in a hibernating (low-drag) state -- to their reconstruction with $305$ POD modes and their reconstruction from the autoencoder with $d_h=25$. When reconstructing the simpler hibernating state the reconstruction from POD and the autoencoder match almost exactly. In the case of the bursting snapshot, both the POD and the autoencoder still match the full high-dimensional state well, accurately capturing the magnitude and location of the streamwise velocity. In the case of POD, some error is introduced as it distorts the details in the DNS, and then the autoencoder further smooths some of these details. However, considering that we reduced the dimension of the problem from $\mathcal{O}(10^5)$ to $25$, the reconstruction is excellent. Now that we can represent the state $s$ in the manifold coordinate system as $h$, the next step is learning a time evolution model to control.

\begin{table}
	\captionsetup{justification=raggedright}
	\caption{{Architectures of NNs. ``Shape" indicates the dimension of each layer, ``Activation" the corresponding activation functions, and ``sig" is the sigmoid activation.``Learning Rate" gives the learning rate for the Adam optimizer \citep{Kingma2015}. When multiple learning rates are noted, the value was changed from one value to the next at even intervals during training.}}
	\centering
        \resizebox{\textwidth}{!}{
		\begin{tabular}{l*{6}{c}r}
			Function & Shape & Activation & Learning Rate \\
			\hline
			$\mathcal{E}$		& 500/1000/$d_h$ \quad           & sig/lin         & $[10^{-3},10^{-4}]$ \\
			$\mathcal{D}$		& $d_h$/1000/500 \quad           & sig/lin         & $[10^{-3},10^{-4}]$ \\
			$g$	    & $d_h$/200/200/200/200/$d_h$ \quad  & sig/sig/sig/sig/lin & $[10^{-2},10^{-3},10^{-4}]$ \\
                $\mathcal{R}$		& $d_h$/100/100/1 \quad          & sig/sig/lin    & $[10^{-3}]$ \\
                $\mathbb{O}$        & 1024/100/100/$d_h$ \quad      & sig/sig/lin       & $[10^{-3}]$ \\
			$\pi$	            & $d_h$/256/128/128/2/1 \quad      & ReLU/ReLU/ReLU/lin/Tanh & $[3\cdot 10^{-4}]$ \\
   			$Q_1$	            & $d_h$/256/128/128/1 \quad      & ReLU/ReLU/ReLU/lin & $[3\cdot 10^{-3}]$ \\
            $Q_2$	            & $d_h$/256/128/128/1 \quad      & ReLU/ReLU/ReLU/lin & $[3\cdot 10^{-3}]$ \\
            $V$	            & $d_h$/256/128/128/1 \quad      & ReLU/ReLU/ReLU/lin & $[3\cdot 10^{-3}]$ \\
		\label{Table1}
		\end{tabular}}
\end{table}

\subsection{DManD Performance} \label{sec:Resultsb}

We now describe the neural ODE training and performance of the resulting model. Before training, we normalize $h$ by subtracting the mean and dividing each component by its standard deviation. In Table \ref{Table1} we show the architecture for the neural ODEs used in this section. 
We trained four neural ODE models for each of the four autoencoders. Then we selected the autoencoder and neural ODE pair that best reconstructed the statistics we report below. Again, we trained the neural ODEs until we no longer saw an improvement in performance.

\begin{figure} 
	\includegraphics[trim=0 0 0 0,width=\textwidth,clip]{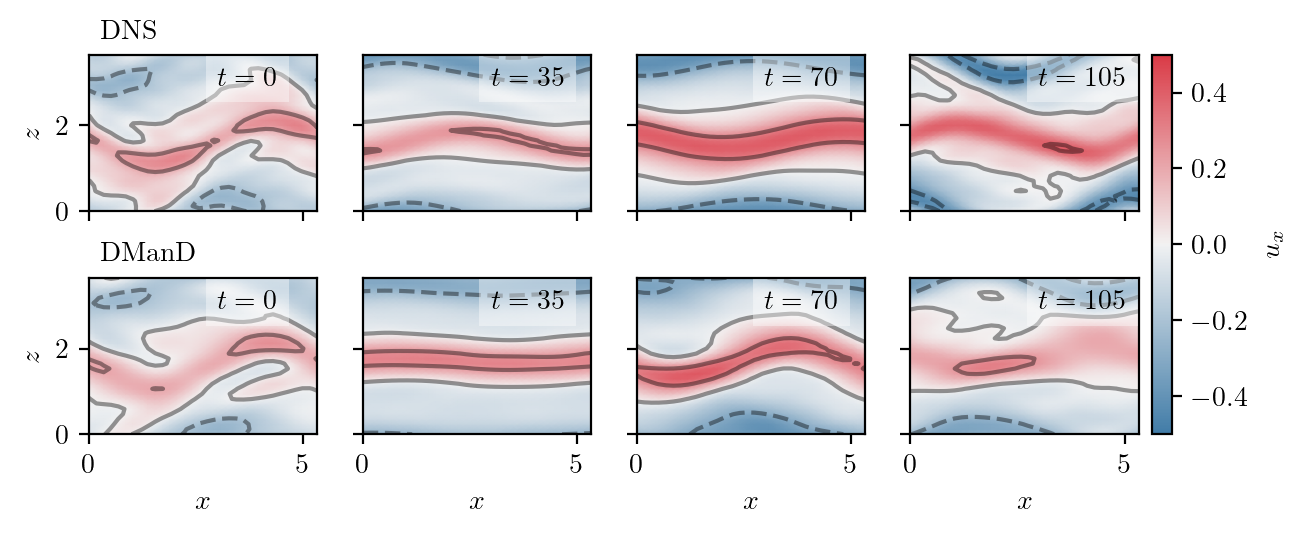}
	\captionsetup[subfigure]{labelformat=empty}
	\caption{Comparison of snapshots from a randomly actuated trajectory from the DNS (top) with the DManD reconstruction (bottom). Times are given at the top right of each image.}
	\label{fig:SSP_Snaps}
\end{figure}

The first statistics we investigate validate the ability of the model to track the true dynamics over short times. In Fig.\ \ref{fig:SSP_Snaps} we show an example of a randomly actuated trajectory from the DNS and reconstructed with the DManD model. In the first two snapshots, the trajectories are in quantitative agreement, after which the trajectories still appear qualitatively quite similar. Both the DNS and the DManD model exhibit the streak breakdown and regeneration cycle over this series of snapshots. At $t=0$ rolls are forming, at $t=35$ these rolls lift low-speed fluid off the wall forming streaks, at $t=70$ the streaks become wavy, and, finally, at $t=105$ the streaks have broken down.

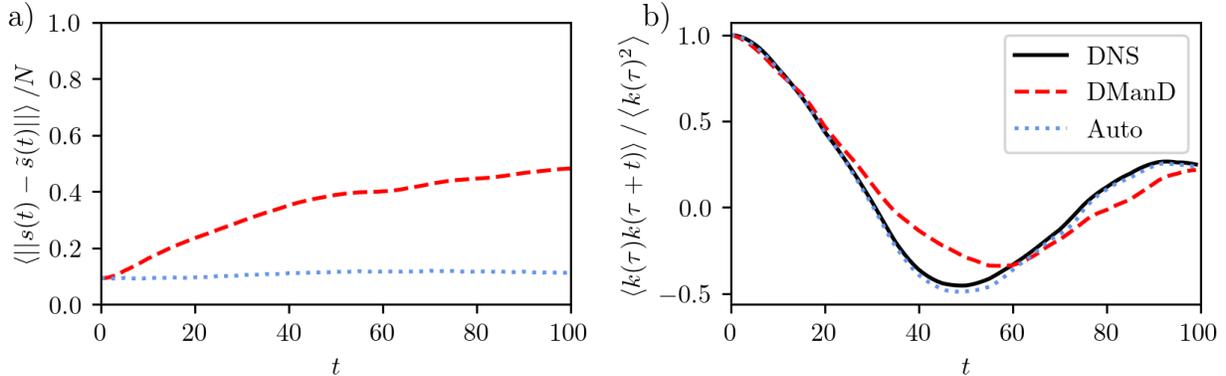
\begin{figure} 
    \centering
	\includegraphics[trim=0 0 0 0,width=\textwidth,clip]{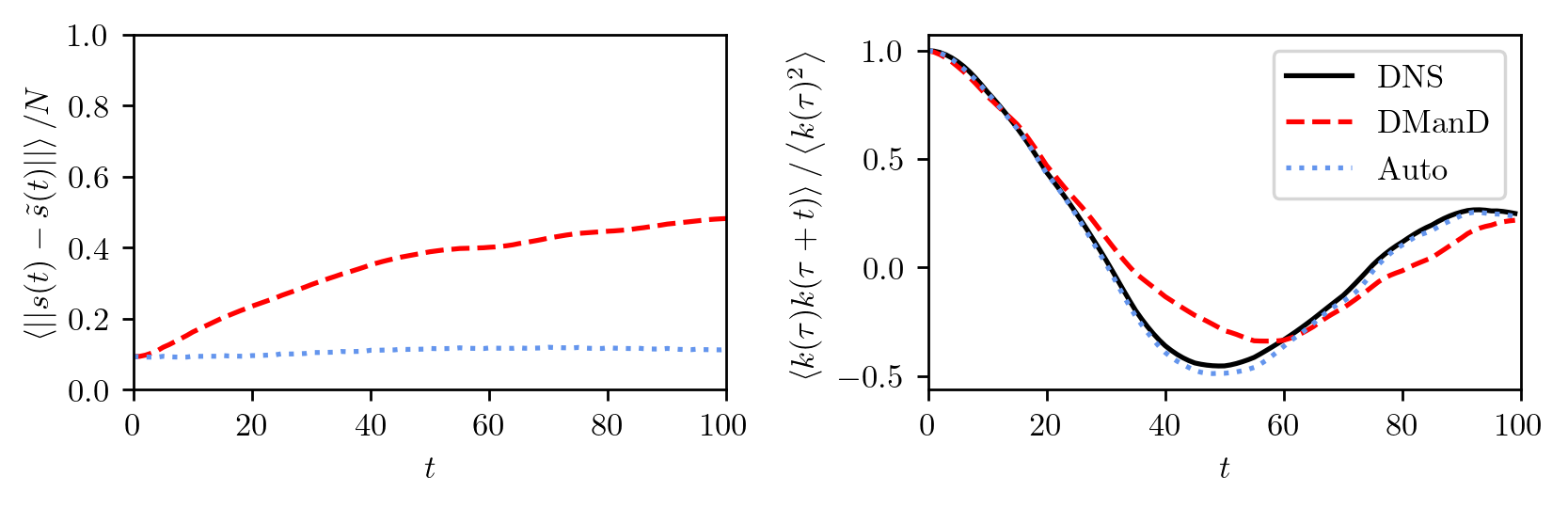}
	\captionsetup[subfigure]{labelformat=empty}
 	\begin{picture}(0,0)
	\put(-230,155){\contour{white}{ \textcolor{black}{a)}}}
	\put(10,155){\contour{white}{ \textcolor{black}{b)}}}
	\end{picture} 
        {\phantomsubcaption\label{fig:Tracking}}
        {\phantomsubcaption\label{fig:Corr}}

	\caption{(a) Ensemble averaged tracking error and (b) temporal autocorrelation of kinetic energy for the DNS, the DManD model, and the autoencoder. ``Auto" represents putting the DNS trajectories through the autoencoder without any time prediction: i.e.~this represents the error of just reducing, then expanding the dimension.}
	\label{fig:TrackingCorrV1}
\end{figure}

Next, we consider model performance when averaged over many trajectories. \ALrevise{In all of these cases we show both the error from time evolving with the DManD model, and the error incurred due to the autoencoder. We compute the error due to the autoencoder by inputting the true DNS solution through the autoencoder and computing the relevant statistic.}
Figure \ref{fig:Tracking} is a plot of the normalized ensemble-averaged tracking error as a function of time. We compute this error by finding the difference between $s(t)$ and $\tilde{s}(t)$ 
from the DManD model for 100 initial conditions. 
Then, we normalize this error by computing the difference between two states on the attractor at random times $t_i$ and $t_j$  $N=\left<||s_{t_i}-s_{t_j}||\right>$. With this normalization, the long-time dynamics of two slightly perturbed initial conditions with different random actuation sequences should approach unity. \MDGrevise{The error at $t=0$ represents the discrepancy between the full state and its reconstruction when passed through the autoencoder.} The ensembled-averaged tracking error rises steadily at one slope for the first $50$ time units and then at a lower slope after that. 
Once the curve levels off, the true and model trajectories have become uncorrelated. Based on these results, the model tracks well for $\sim 50$ time units. For reference, the Lyapunov time (inverse of the Lyapunov exponent) for the unactuated system is $\tau_L=48$ time units \cite{Inubushi2015}. 

In addition to the tracking error, we also check the ability of the DManD model to capture the temporal autocorrelation of the kinetic energy. 
To compute this autocorrelation we take the instantaneous kinetic energy of the flow
\begin{equation} \label{eq:KE}
	 E(t)=\frac{1}{2L_x L_z} \int_0^{L_z} \int_{-1}^1 \int_0^{L_x} \frac{1}{2}\mathbf{u}\cdot\mathbf{u} \;d\boldsymbol{x},
	 \end{equation}
	 and subtract the mean to yield $k(t)=E(t)-\left<E\right>$. Figure \ref{fig:Corr} shows this temporal autocorrelation computed from \ALrevise{the DNS, the DManD model, and the autoencoder.} We see the temporal autocorrelation of the model matches the true temporal autocorrelation closely over the first $\sim 30$ time units. 

\begin{figure} 
    \centering
	\includegraphics[trim=0 0 0 0,width=\textwidth,clip]{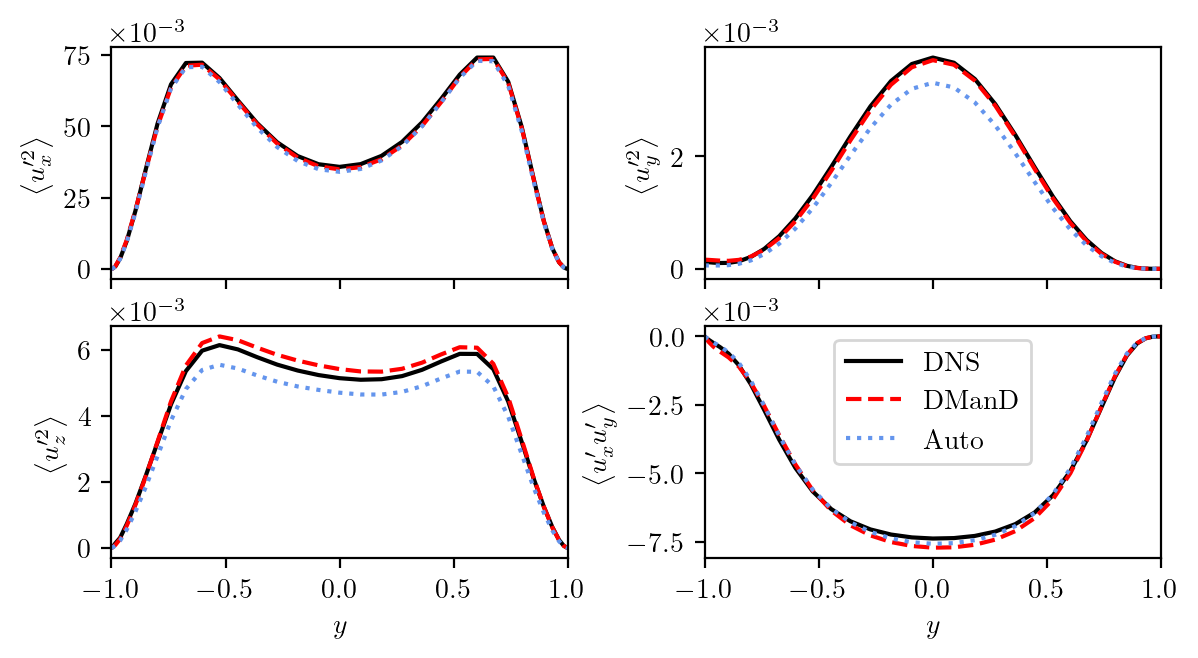}
	\captionsetup[subfigure]{labelformat=empty}
	\caption{Four components of the Reynolds stress for the DNS, the DManD model, and the autoencoder.}
	\label{fig:StressV1}
\end{figure}

Due to the chaotic nature of this system, short-time tracking is limited by the Lyapunov time.  Nevertheless, a good model should still be able to capture the long-time statistics of the original system.
 In Fig.\ \ref{fig:StressV1} we show the four components of the Reynolds stress for the DNS, 
 \ALrevise{the DManD model, and the autoencoder.}
We computed these statistics by averaging over the entire testing dataset (100 trajectories 300 time units in length).
For all these quantities, the DManD model is in excellent agreement with the DNS. In the cases of $\left<u^2_y\right>$ and $\left<u^2_z\right>$, surprisingly, the autoencoder appears to perform worse than the DManD model at matching the DNS. With perfect prediction of the DManD model, it would exactly match the autoencoder. This slight disagreement indicates that some of the states in the manifold coordinates ($h$) are driven somewhat outside the expected range of values.

\begin{figure} 
    \centering
	\includegraphics[trim=0 0 0 0,width=\textwidth,clip]{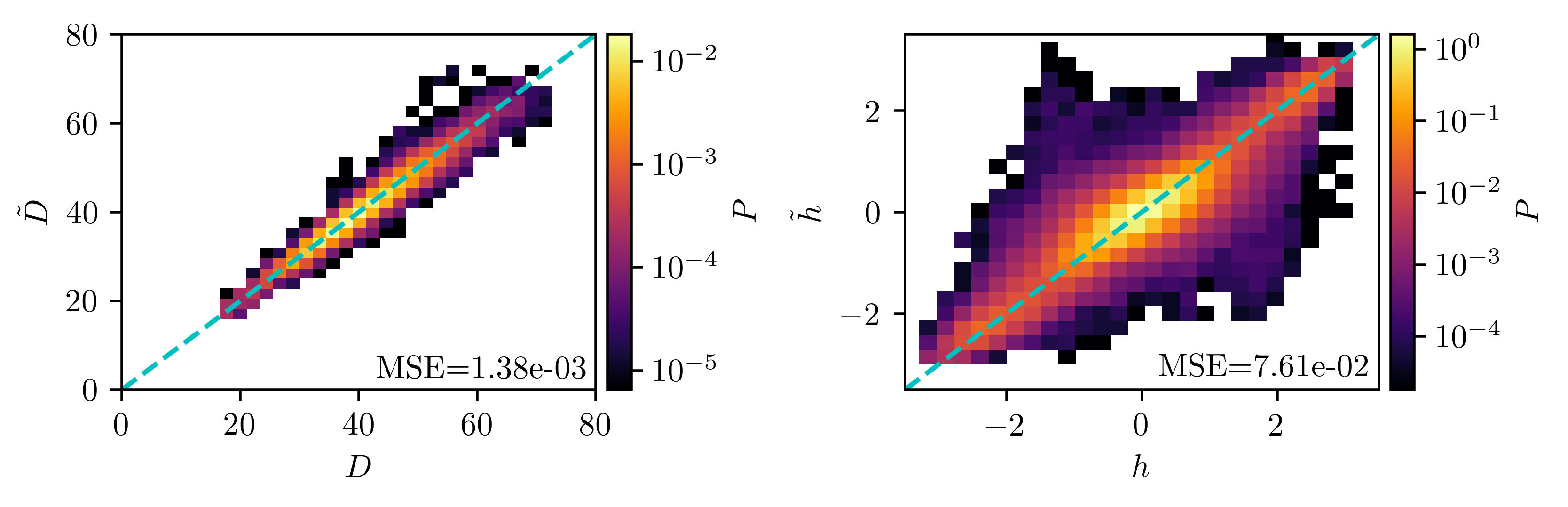}
	\captionsetup[subfigure]{labelformat=empty}
  	\begin{picture}(0,0)
	\put(-230,150){\contour{white}{ \textcolor{black}{a)}}}
	\put(15,150){\contour{white}{ \textcolor{black}{b)}}}
	\end{picture} 
        {\phantomsubcaption\label{fig:ParityPDF-a}}
        {\phantomsubcaption\label{fig:ParityPDF-b}}
	\caption{(a) Joint PDF of the true ($D$) and predicted drag ($\tilde{D}$) from the reward network, Eq.~\ref{eq:rewardnetwork}. (b) Joint PDF of the true ($h$) and predicted ($\tilde{h}$) low-dimensional state from the observation network, Eq.~\ref{eq:hfromwallstress}. Note the logarithmic scales. The cyan line indicates perfect reconstruction, and the MSE is for $D$ and $h$ with the mean subtracted and divided by the standard deviation.}
	\label{fig:ParityPDF}
\end{figure}
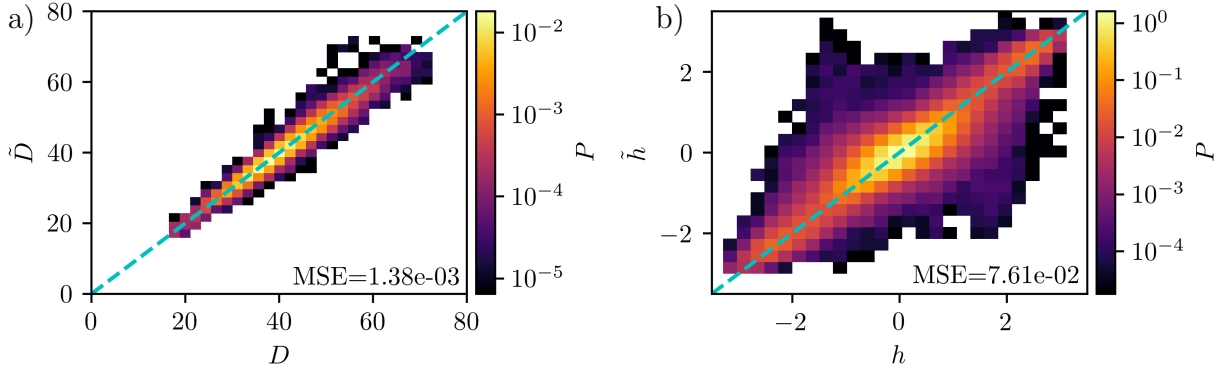

All of the statistics shown so far indicate that the DManD model accurately captures the dynamics of the randomly actuated DNS with only $d_h=25$ degrees of freedom.  The last step before using this model in the RL framework is training the reward network as described in Sec.\ \ref{sec:FrameworkModeling}. We trained a NN with the architecture in Table \ref{Table1}. In Fig.\ \ref{fig:ParityPDF-a} we show a PDF of the parity plot between the true and predicted drag on test data, and report the MSE on the normalized data. The excellent agreement indicates we can compute accurate values of the reward directly from $h$.


Finally, although we computed the manifold coordinate system directly from the state, a natural extension of this work is to use a more limited set of observations. As a first step in this direction, we trained a NN $\mathbb{O}(\cdot)$ to map $32 \times 32$ wall shear rate observations at the bottom wall to the manifold coordinate system:
\begin{equation}
    \tilde{h}=\mathbb{O}(\partial_y u_x|_{y=-1};\theta_O).\label{eq:hfromwallstress}
\end{equation} The details of this NN are included in Table \ref{Table1}. This mapping allows us to test if it is possible to directly use wall observables that are experimentally realizable with our control policy. In Fig.\ \ref{fig:ParityPDF-b} we show the parity plot of reconstructing $h$ from wall observables and report the MSE for normalized data. \MDGrevise{While the parity plot is not as sharply peaked as we might like, note that it is shown on a logarithmic scale.}

\subsection{DManD-RL Performance} \label{sec:Resultsc}
Now with an efficient and low-dimensional model of the underlying dynamics of the turbulent flow in hand, we can quickly obtain a control agent by performing deep RL on the DManD model rather than the original costly DNS. In this work, we employ the Soft Actor-Critic (SAC) RL algorithm \citep{SAC}, which requires training a policy function (i.e.\ agent) $\pi$, two critic functions, $Q_1$ and $Q_2$, and a value function $V$. 
The RL networks are trained for 10,000 episodes, with each episode consisting of DManD model trajectories of 300 time units. The initial condition for each episode is selected at random from on-attractor turbulent states. Here we choose the action time to be $\tau=5.0$ and $\gamma=0.99$. \MDGrevise{Accordingly, the discount factor over one Lyapunov time $\tau_L\approx \KZrevise{48}$  of the unactuated system is $\gamma^{\tau_L/\tau}\approx 0.9$.}

Once trained, we apply the DManD-RL agent to the original turbulent DNS, with which it has never directly seen or interacted. To deploy the agent for control, we insert the already-trained encoder, $\chi$, between the agent and the environment to map state observations of the turbulent DNS to the manifold representation, $h$, as this is the observation space where the DManD-RL agent was trained.

\begin{figure} 
    \centering
	\includegraphics[trim=0 0 0 0,width=\textwidth,clip]{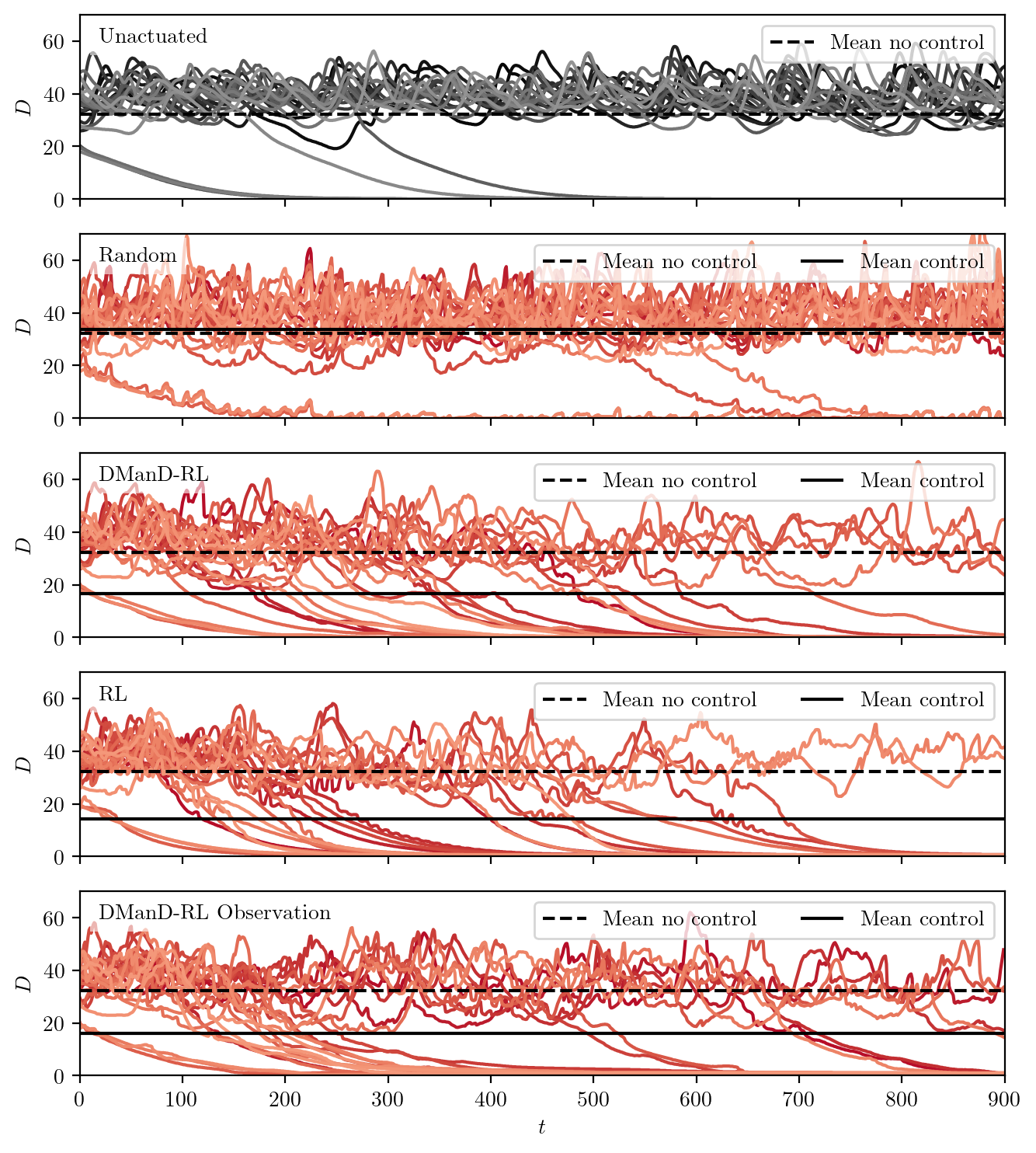}
	\captionsetup[subfigure]{labelformat=empty}
 	\begin{picture}(0,0)
	\put(-225,520){\contour{white}{ \textcolor{black}{a)}}}
	\put(-225,421){\contour{white}{ \textcolor{black}{b)}}}
 	\put(-225,322){\contour{white}{ \textcolor{black}{c)}}}
	\put(-225,223){\contour{white}{ \textcolor{black}{d)}}}
 	\put(-225,124){\contour{white}{ \textcolor{black}{e)}}}
	\end{picture} 
	\begin{subfigure}[b]{0\textwidth}\caption{}\vspace{-10mm}\label{fig:DragAll-a}\end{subfigure}
	\begin{subfigure}[b]{0\textwidth}\caption{}\vspace{-10mm}\label{fig:DragAll-b}\end{subfigure}
	\begin{subfigure}[b]{0\textwidth}\caption{}\vspace{-10mm}\label{fig:DragAll-c}\end{subfigure}
	\begin{subfigure}[b]{0\textwidth}\caption{}\vspace{-10mm}\label{fig:DragAll-d}\end{subfigure}
	\begin{subfigure}[b]{0\textwidth}\caption{}\vspace{-10mm}\label{fig:DragAll-e}\end{subfigure} 
	\caption{Trajectories beginning from test initial conditions with (a) no control, (b) random actuations, (c) DManD-RL control, (d) DNS-based RL control, and (e) DManD-RL control using wall observations. Each figure shows 25 test trajectories.} 
	\label{fig:DragAll}
\end{figure} 

Shown in Fig.\ \ref{fig:DragAll} are times series, of length 900, of the drag in various cases from DNS trials using 25 different unseen initial conditions. \KZrevise{Here we define the \% drag reduction, \begin{equation}
    DR=\dfrac{D_{0}-D}{D_0}\cdot 100\%,
    \label{eq:dr}
\end{equation}
where $D_0$, and $D$ is the time averaged drag experienced for the 25 test trajectories over 900 time units under no control and control, respectively.}
In Fig.\ \ref{fig:DragAll-a} we plot the drag of the 25 test trajectories in the absence of control and highlight that the system predominately remains turbulent, with a few laminarizations. In Fig.\ \ref{fig:DragAll-b} we show that utilizing a random jet actuation policy does not reduce drag and even \KZrevise{results in increased drag $DR=-4\%$} across the 25 trajectories over the 900 time unit window. In Fig.\ \ref{fig:DragAll-c} we show that our DManD-RL agent is able to significantly reduce the drag of the turbulent DNS and laminarize 21/25 turbulent initial conditions\KZrevise{, yielding $DR=48\%$}. In comparison, we show in Fig.\ \ref{fig:DragAll-d} that the conventionally trained RL agent, which was trained directly on the turbulent DNS, results \KZrevise{laminarization of 23/25 test trajectories with $DR=56\%$, similar to that of} the DManD-RL method. We highlight here, however, that the DManD-RL control agent was obtained at a small fraction of the computational cost compared to its conventional  counterpart. For reference, the DManD-RL agent required $\sim 3.7$s per training episode, while a conventional application of deep RL required $\sim 1630$s per episode, on a 2.40GHz Intel Xeon CPU E5-2640 v4. This corresponds to a 440 times speedup in training time.

Finally, we now limit the state observation of the DManD-RL agent to information observable at the wall i.e.\ wall shear rate, by pairing the DManD-RL agent trained for Fig.\ \ref{fig:DragAll-c} with the observation network ($\mathbb{O}(\partial_y u_x|_{y=-1};\theta_O)$). We demonstrate in Fig.\ \ref{fig:DragAll-e} that the agent with only access to wall observations, $a_t=\pi(\mathbb{O}(\partial_y u_x|_{y=-1};\theta_O);\theta_A)$, performs just as well as its counterparts, \KZrevise{with 20/25 test trajectories laminarizing and  $DR=50\%$ over the 900 time unit window.} \KZrevise{Here we note that $DR$ is influenced by how many and how quickly test trajectories laminarize, as well as how turbulent transient trajectories are attenuated by the various controllers and we emphasize that all of the RL controllers perform similarly.}

\subsubsection{Interpretation of the mechanism of drag reduction}

\begin{figure} 
    \centering
	\includegraphics[trim=0 0 0 5,width=\textwidth,clip]{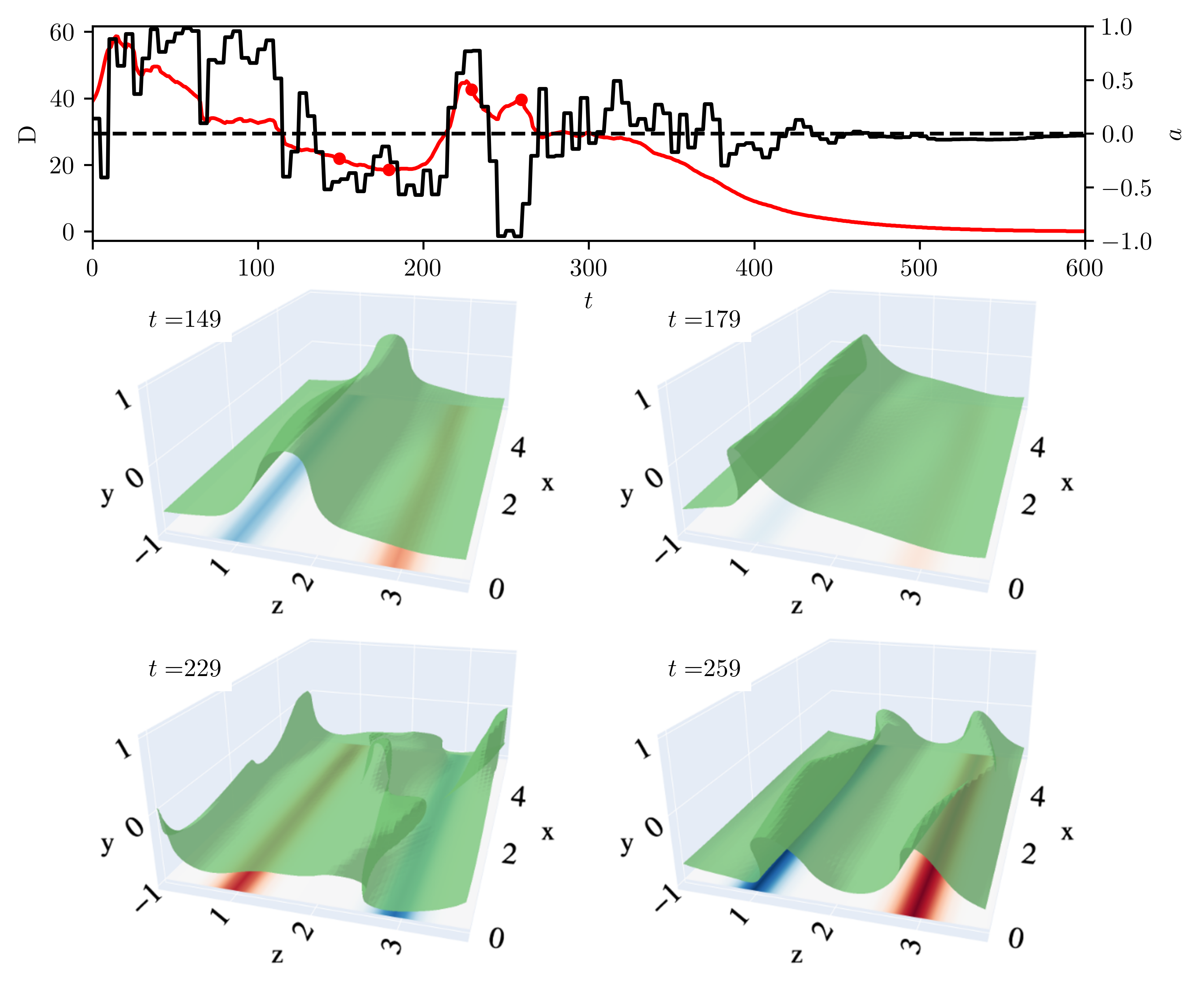}
	\captionsetup[subfigure]{labelformat=empty}
 	\begin{picture}(0,0)
	\put(-225,385){\contour{white}{ \textcolor{black}{a)}}}
	\put(-185,260){\contour{white}{ \textcolor{black}{b)}}}
 	\put(17,260){\contour{white}{ \textcolor{black}{c)}}}
	\put(-185,125){\contour{white}{ \textcolor{black}{d)}}}
 	\put(17,125){\contour{white}{ \textcolor{black}{e)}}}
	\end{picture} 
 	\begin{subfigure}[b]{0\textwidth}\caption{}\vspace{-10mm}\label{fig:Mechanism1-a}\end{subfigure}
	\begin{subfigure}[b]{0\textwidth}\caption{}\vspace{-10mm}\label{fig:Mechanism1-b}\end{subfigure}
	\begin{subfigure}[b]{0\textwidth}\caption{}\vspace{-10mm}\label{fig:Mechanism1-c}\end{subfigure}
	\begin{subfigure}[b]{0\textwidth}\caption{}\vspace{-10mm}\label{fig:Mechanism1-d}\end{subfigure}
	\begin{subfigure}[b]{0\textwidth}\caption{}\vspace{-10mm}\label{fig:Mechanism1-e}\end{subfigure} 
 	\vspace{-5mm}
	\caption{(a) Time series of drag (red) and actuation signal of the left jet (black) for an example DNS trajectory controlled by the DManD-RL agent. (b)-(e) snapshots of the trajectory at times marked in (a). These snapshots include an isosurface of $u_x=-0.35$ and the jet actutations (red is fluid injection, blue is fluid suction).}
	\label{fig:Mechanism1}
 	\vspace{-5mm}
\end{figure}
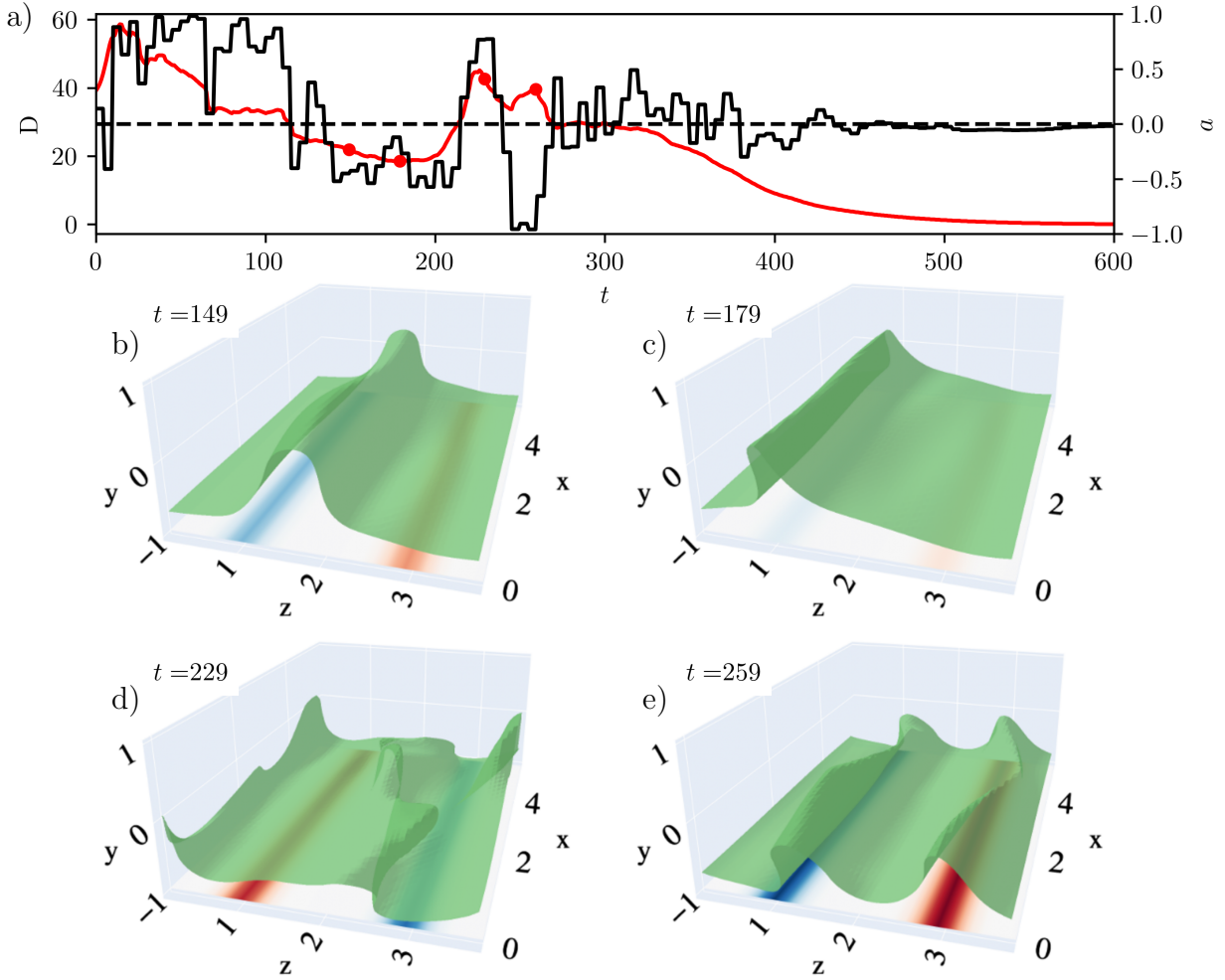

\MDGrevise{Given the observed effectiveness of the control policy discovered by the RL algorithm, it is desirable to understand how the controller is modifying the flow. In this section we describe two sets of observations that may shed some light on this issue. The first focuses on the control action in the time preceding a laminarization event.}   
Shown in Fig.\ \ref{fig:Mechanism1-a} is a time series depicting the drag and left actuator control signal of a DNS controlled by the DManD-RL agent. \MDGrevise{Fig.\ \ref{fig:Mechanism1-b} shows the wall actuation and an isosurface of streamwise velocity illustrating the streamwise streak structure at $t=149$, indicated by the first red dot on the drag time series in {fig:Mechanism1-a}. Here the left slot jet is sucking, and the right blowing, drawing the fluid (and low-speed streak indicated by the isosurface) to the left. At $t=179$ (second red dot and Fig.\ \ref{fig:Mechanism1-c}), the streak is now located over the left jet. }
Once the streak is above the left jet, the agent executes a series of actions that destabilize the low-speed streak, causing it to break down, shown in Fig.\ \ref{fig:Mechanism1-d}. In the wake of the collapse, the agent initiates a strong actuation that leads to the formation of \emph{two} low-speed streaks, shown in Fig.\ \ref{fig:Mechanism1-e}. This double low-speed streak structure then proceeds to decay to the laminar state, with the agent applying weak, attenuating control actions to expedite the process. 

A DNS initialized with this double-streak structure flow field in the absence of control results in the natural laminarization of the flow. In wall-bounded turbulence, streaks take on a characteristic spacing of 100 wall units \cite{Smith1983}, which is approximately the width of the MFU cell \cite{Hamilton1995,Jimenez1991}. The two-streak state has a spanwise length scale that is too small and thus too dissipative to self-sustain, leading to breakdown in the SSP and laminarization of the flow. In approximately half of the trials that laminarized, this double-streak structure appeared before laminarization. 

\MDGrevise{This is a very interesting and counterintuitive strategy, particularly because it is intrinsically \emph{nonlinear}. At any given instant, the two-jet system can only drive a wall-normal flow with the same fundamental wavelength as the domain. But here the RL agent implements a time-dependent policy that ends up generating a flow structure with, roughly speaking, half the wavelength of the domain. I.e.~it has figured out a way to drive structure to a smaller scale, where viscosity can take over and drive the flow to laminar. } 




\MDGrevise{The second set of observations we highlight addresses the relationship between the RL control policy and a flow control strategy, widely-studied with simulations, that we mentioned in Sec.\ \ref{sec:Introduction} -- opposition control. }
In opposition control, the entire wall-normal velocity field at both walls is set to have the opposite sign as the wall-normal velocity at an $x-z$ ``detection plane"  located at some $y$ near the wall.
We highlight that this method possesses much greater control authority than in the present work, as it has full spatial control at both walls whereas here we only actuate two spatially localized slot jets on a single wall. 
Furthermore, our controller holds a constant actuation for multiple time units, whereas opposition control updates on the time scale of a time step.
With these differences in mind, we investigate trajectories controlled by the DManD-RL policy in the context of potential similarities to opposition control. To make this comparison, we first must characterize the wall-normal velocity at the detection plane, but because the jets span the length of the channel this velocity varies. 
\MDGrevise{We take a characteristic detection-plane velocity to be the wall-normal velocity averaged over a jet weighted by the shape of the jet, denoted $\left< u_y\right>_J$.}

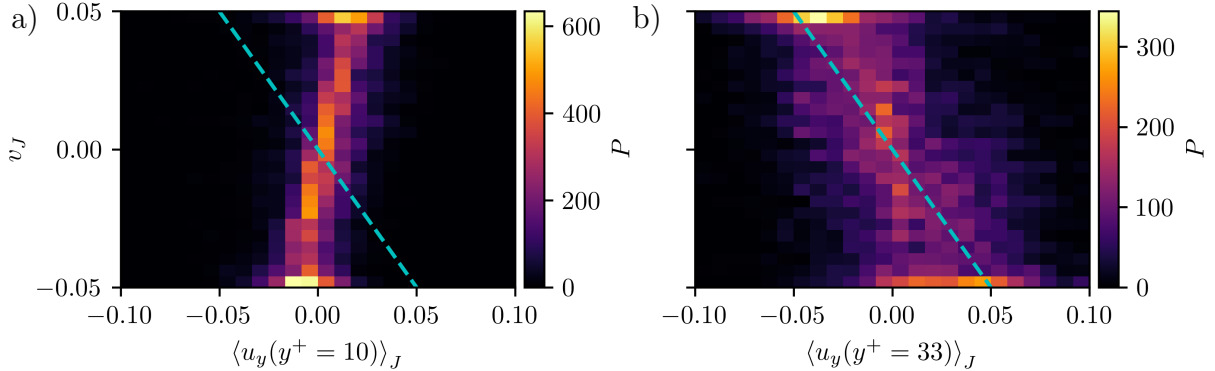
\begin{figure} 
    \centering
	\includegraphics[trim=0 0 0 0,width=\textwidth,clip]{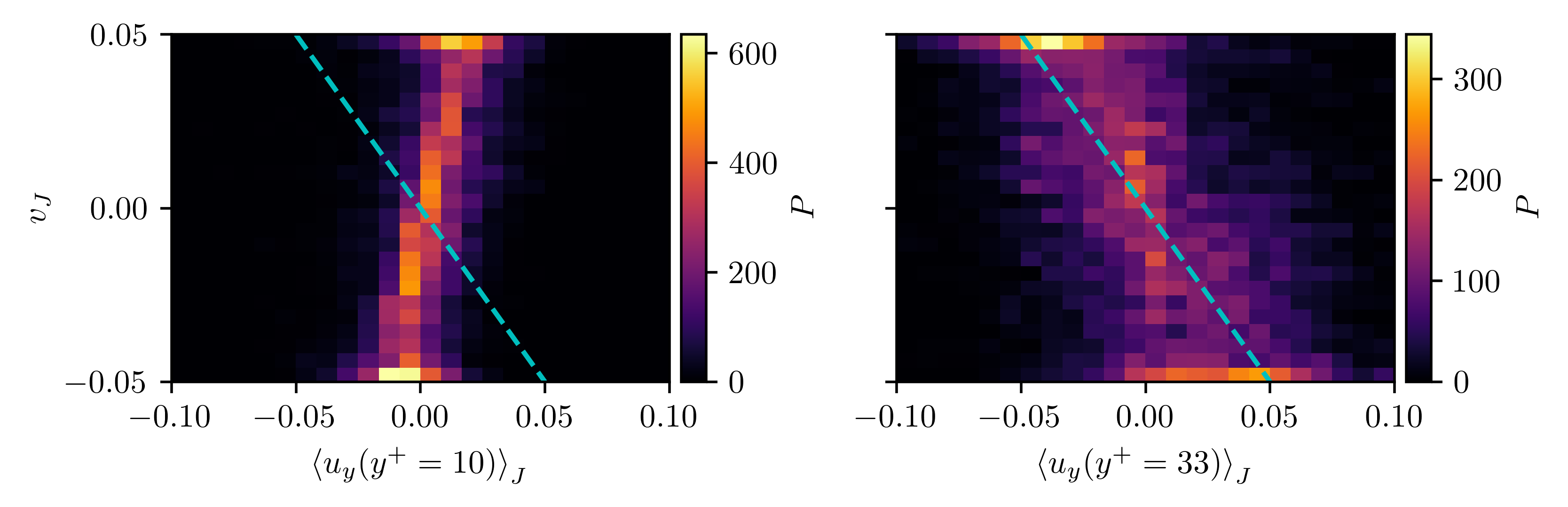}
	\captionsetup[subfigure]{labelformat=empty}
  	\begin{picture}(0,0)
	\put(-225,150){\contour{white}{ \textcolor{black}{a)}}}
	\put(10,150){\contour{white}{ \textcolor{black}{b)}}}
	\end{picture} 
        {\phantomsubcaption\label{fig:PDFControl-a}}
        {\phantomsubcaption\label{fig:PDFControl-b}}
        
	\caption{Joint PDF of max jet velocity and average wall-normal velocity at (a) $y^+\approx10$ and (b) $y^+=33$ (the cyan dotted line is $v_J=-\left< u_y\right>_J$).}
	\label{fig:PDFControl}
\end{figure}

In Fig.\ \ref{fig:PDFControl} we plot a PDF of the maximum jet velocity $v_J$ 
and \ALrevise{the wall-normal velocity} $\left< u_y\right>_J$ \ALrevise{(for both jets separately)} at a detection plane of $y^+\approx10$ (Fig.\ \ref{fig:PDFControl-a}) and $y^+=33$ (Fig.\ \ref{fig:PDFControl-b}).
\ALrevise{We remove laminarization events from this dataset by omitting states with a drag less than 10 because we are interested in the control behavior while the flow is still turbulent.} 
\ALrevise{In these plots, we denote opposition control with a unit gain ($v_J=-\left< u_y\right>_J$) with the cyan line. If the agent performs opposition control, the joint PDF should show a tight distribution with a negative slope.}
At the detection plane $y^+\approx 10$ that has been reported to be optimal for opposition control, \cite{Choi1994} the RL agent behaves like an anti-opposition controller \ALrevise{because there is a positive slope in the PDF}.
However, at a higher detection plane of $y^+=33$, the control agent has a high probability of actuating with sign opposite to $\left< u_y\right>_J$, as in opposition control. We highlight that the PDF is quite broad, which indicates although the control is opposition-like, the control response is much more complex and diverse than that of opposition control.

Furthermore, we can directly compare our results to \citet{Ibrahim2019}, who applied standard opposition control to this same Couette domain. In their work, the authors selected a detection plane of $y^+=10$ and varied the wall-normal velocity scale $\phi=[0.1,0.2,0.5,1]$. For these parameters, the authors found that the probability of the flow remaining turbulent after $900$ time units of control, the length of our test trajectories shown in Fig.\ \ref{fig:DragAll}, to be approximately $[0.77,0.72,0.69,0.42]$. For comparison, with $21/25$ trajectories laminarizing, our DManD-RL agent's probability of turbulence is $0.16$. This improvement is extremely promising for the future use of RL in controlling turbulent flows, especially as the DManD-RL agent had major \MDGrevise{and realistic} restrictions placed on its control authority when compared to opposition control. We also note that this improvement in control further supports the claim that the agent learned a much more complex and diverse control strategy than opposition control. 

\color{black}


\section{Conclusions} \label{sec:Conclusions}
In this paper, we efficiently obtained a control strategy from a limited data set to reduce the drag in a turbulent Couette flow DNS via the control of two streamwise slot jets using the DManD-RL framework.
Using a combination of POD and autoencoders, we extracted a low-dimensional manifold representation of the data, whose dynamics we modeled using a neural ODE. We show that our 25-dimensional DManD model qualitatively captures the turbulent self-sustaining process and has good short-time predictive capabilities, matching the kinetic energy temporal autocorrelation for 30 time units.
Furthermore, our DManD model excellently captures long-time statistics such as Reynolds stress. In order for this DManD model to be viable for RL, we added an additional reward network to predict the system drag, $D$, given the low-dimensional manifold state, $h$.

We then obtained a control strategy from our DManD model using deep RL, which successfully transferred to and controlled the original DNS. We were able to expeditiously train an RL agent using DManD-RL 440 times faster than a direct application of deep RL to the DNS.

We additionally emphasize here that the data generation and DManD model training in this framework is a fixed one-time cost that is greatly exceeded by the cost of conventional DNS-based RL training. For reference, we found the DNS-based RL training required over a month of training to accomplish 1,000 training episodes, while all steps of the DManD-RL framework (data generation, dynamics model training, 10,000 episodes of RL training) were accomplished within two days.

We find that our DManD-RL agent can consistently drive unseen turbulent initial conditions in the original DNS to the laminar state, despite never having any direct observations or interactions with the DNS. We also demonstrate that there exists a mapping between wall observables and the manifold state, which allowed us to apply the DManD-RL agent with equal effectiveness using only wall shear rate observations.

When investigating the mechanistic nature of the learned control strategy, we observed multiple control strategies executed by the DManD-RL agent. One novel strategy the agent appears to employ consists of manipulating the low-speed streak to a preferred location, causing the breakdown of the streak, and in the wake of the break-down forming two low-speed streaks in its place. These two low-speed streaks are unsustainable within the domain, breaking the SSP and resulting in laminarization.

When comparing the ensemble behavior of our control agent to opposition control, we find that the controller behaves anti-oppositionally at the \MDGrevise{commonly-used} detection plane location of $y^+=10$. At a detection plane of $y^+=33$, we find that the ensemble behavior is opposition-like, however, the broadness of the control action distribution\MDGrevise{, as well as the observation of the two-streak structure,} leads us to conclude that the learned controller behavior is much more complex and diverse than a simple opposition feedback rule. We also compare our DManD-RL agent's control performance to that of opposition control and we find that our control agent out-performs opposition control by a notable margin (16\% vs. 42\% probability of remaining turbulent after 900 time units of control) despite our control set-up and agent possessing much greater restrictions on it spatial and temporal control authority (fixed time intervals of control and localized jets on only the lower wall) compared to opposition control.

\section*{Acknowledgments}
\noindent This work was supported by AFOSR FA9550-18-1-0174 and ONR N00014-18-1-2865 (Vannevar Bush Faculty Fellowship).

\bibliographystyle{elsarticle-num-names}
\bibliography{library.bib,jfm.bib}

\end{document}